\pdfoutput=1
\documentclass[aps,prl,twocolumn,a4paper]{revtex4}
\usepackage{graphicx}
\usepackage{amssymb}
\usepackage{amsmath}
\usepackage{color}
\usepackage{dsfont}
\usepackage{paralist}
\usepackage{bm}
\usepackage{verbatim}
\usepackage{textcomp}

\usepackage{refcount}

\def\cL{{\cal L}}
\def\cO{{\cal O}}

\def\bk{{\bf k}}

\def\bq{{\bf q}}
\def\bp{{\bf p}}
\def\br{{\bf r}}

\def\hbsigma{\hat{\boldsymbol \sigma}}

\def\holOne{\mathds{1}}

\begin{document}

\title{Multifractality at non-Anderson disorder-driven transitions
\\
in Weyl semimetals and other systems}


\author{S.V.~Syzranov$^{1,2}$, V.~Gurarie$^{1,2}$, L.~Radzihovsky$^{1,2,3}$}

\affiliation{
$^1$Physics Department, University of Colorado, Boulder, Colorado 80309, USA
\\
$^2$Centre for Theory of Quantum Matter, University of Colorado, Boulder, Colorado 80309, USA
\\
$^3$JILA, NIST, University of Colorado, Boulder, Colorado 80309, USA
}

\begin{abstract}
  {
  Systems with the power-law
  quasiparticle dispersion $\epsilon_\bk\propto k^\alpha$ exhibit non-Anderson disorder-driven
  transitions in dimensions $d>2\alpha$, as exemplified by Weyl semimetals, 1D and 2D arrays
  of ultracold ions with long-range interactions, quantum kicked rotors and semiconductor models in high dimensions.
  We study the wavefunction structure
  in such systems and demonstrate that at these transitions they exhibit fractal behaviour with an infinite
  set of multifractal exponents. The multifractality persists even when the wavefunction
  localisation is forbidden by symmetry or topology and occurs as a result of elastic scattering
  between all momentum states in the band on length scales shorter than the mean free path.
  We calculate explicitly the multifractal spectra
  in semiconductors and Weyl semimetals using one-loop and two-loop renormalisation-group approaches
  slightly above the marginal dimension $d=2\alpha$.  
  }
\end{abstract}



\date{\today}
\maketitle


After half a century of studies, disorder-driven transitions in conducting materials still
motivate extensive research efforts. 
Anderson localisation (AL) transition is
responsible for turning a metal into an insulator when increasing the disorder strength
in dimensions $d\geq2$ and was believed for
several decades to be the only possible disorder-driven transition in non-interacting systems.
AL continues to fascinate researchers by its peculiar and universal properties, such as, e.g.,
{\it multifractality}-- fractal
behaviour of the wavefunctions at the transition with
an infinite set of multifractal exponents\cite{Efetov:book,MirlinEvers}.
\phantom{\cite{Syzranov:Weyl,Syzranov:unconv,Suslov:rare}}

A broad class of systems with the power-law quasiparticle dispersion
$\epsilon_\bk\propto k^\alpha$ in dimensions $d>2\alpha$ displays, however, another single-particle
disorder-driven transition distinct from AL\cite{Syzranov:Weyl,Syzranov:unconv}.
This transition, unlike AL, occurs only
near a band edge\footnote{ 
Here we neglect the Lifshitz tails, strongly suppressed at all energies
in high dimensions\cite{Suslov:rare,Syzranov:unconv}.}
or at
a nodal point (in a semimetal).
It reflects in the critical
behaviour of the disorder-averaged density of states (in contrast with AL),
as well as in other physical observables, e.g., conductivity.

Such a transition
has first been proposed\cite{Fradkin1,Fradkin2} for the specific case of Dirac semimetals ($\alpha=1$, $d=3$)
and has recently sparked vigorous
studies\cite{ShindouMurakami,RyuNomura,Goswami:TIRG,Herbut,OminatoKoshino}\cite{Syzranov:Weyl,Syzranov:unconv}
\cite{Brouwer:WSMcond,Pixley:twotrans,Pixley:ExactZpubliahed,LiuOhtsuki:LateNumerics,Brouwer:exponents,
Shapourian:PhaseDiagr,Bera:Weyl,Syzranov:twoloopZ,PixleyHuse:rare}
of its critical properties in 3D Weyl and Dirac systems\cite{ZHasan:TaAs,ZHasan:TaAs2,Weng:PhotCrystWSM}.
Other playgrounds for the observation of this non-Anderson
disorder-driven transition are 1D and 2D arrays of trapped ultracold 
ions with long-range interactions\cite{Garttner:longrange},
quantum kicked rotors\cite{Syzranov:unconv}
(mappable onto high-dimensional semiconductors), and numerical simulations of Schroedinger
equation in $d\geq5$ dimensions\cite{Markos:review,GarciaGarcia,Slevin,Zharekeshev:4D}.

Despite these comprehensive studies, the wavefunction structure at these non-Anderson disorder-driven
transitions is rather poorly understood. Such transitions are not necessarily accompanied by localisation;
they can occur between two phases of localised states [like in 1D (non-chiral)
chains of trapped ions\cite{Garttner:longrange}]
or between two phases of delocalised states [e.g., in single-node Weyl semimetals (WSMs)]
or between localised and delocalised states (in a high-dimensional semiconductor\cite{Syzranov:unconv}).
Particle wavefunctions in all of these cases are characterised by a correlation length that diverges from
both sides of the transition.

{\it Results.}
In this paper we study microscopically wavefunctions $\psi(\br)$ at the non-Anderson disorder-driven transitions
and demonstrate their {\it multifractal} nature. When delocalised states are allowed by symmetry, dimensions,
and topology, the typical wavefunctions 
at the critical disorder strength 
have a fractal structure and
are characterised by a universal non-linear multifractal spectrum $\Delta_q$, defined\cite{MirlinEvers,Efetov:book}
by the inverse participation ratios (IPRs) 
\begin{align}
	P_q=\int |\psi(\br)|^{2q}d\br
\propto L^{-d(q-1)-\Delta_q}
\end{align}
in the limit of an infinite system size $L\rightarrow\infty$.
Such multifractal behaviour persists even if the wavefunctions are delocalised on both sides of the
transition (like in a single-node WSM). Unlike the AL transition, here the multifractal spectrum $\Delta_q$
is determined by the elastic scattering on length scales shorter than the mean free path.
In systems that allow for localised states near the transition, 
these states scale as $\int |\psi(\br)|^{2q}d\br \propto \xi^{-d(q-1)-\Delta_q}$ when approaching it,
where $\xi$ is the localisation length divergent at the transition.
In this paper we also calculate the multifractal spectrum $\Delta_q$ explicitly for several systems.

We find the multifractal spectrum of a
disordered system with the power-law quasipatricle dispersion 
$\epsilon_\bk = a |\bk|^\alpha$ in dimensions $d>2\alpha$ in the orthogonal symmetry class,
in the expansion in powers of $\varepsilon$, to be
\begin{equation}
	\Delta_q^{\text{Semicond}}=\frac{1}{2}\varepsilon q(q-1)+{\cal O}(\varepsilon^2),
	\label{SemicondResult}
\end{equation}
where $\varepsilon=2\alpha-d$ (and $\varepsilon<0$).

For Weyl semimetals, $\epsilon_\bk=\hbsigma\bk$, in $d=2-\varepsilon$ dimensions 
\begin{equation}
	\Delta_q^{\text{WSM}}=-\frac{3}{8}\varepsilon^2 q(q-1)+{\cal O}(\varepsilon^3).
	\label{WeylResult}
\end{equation}
{We note that on
sufficiently short length scales the multifractality in a Weyl semimetal
is similar (with $\varepsilon$ replaced by the dimensionless disorder strength)
to that of 2D Dirac fermions on the surfaces of a 3D topological insulator, studied in Ref.~\cite{Foster:multifract2D}.
Although there is no phase transition in such 2D systems, their wavefunctions display non-universal short-length
multifractal behaviour\cite{Foster:multifract2D,Foster:multifract2Dnumerics}, in contrast with the universal
multifractality (\ref{WeylResult}) that persists at all length scales.}

\begin{figure}[t]
	\centering
	\includegraphics[width=0.4\textwidth]{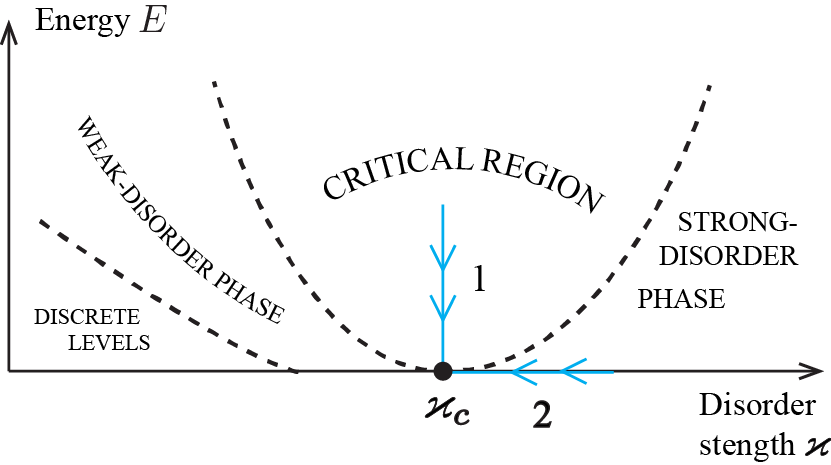}
	\caption{(Colour online) Phase diagram of a finite-size system (with $|\varepsilon|\gtrsim 1$)
	near a non-Anderson disorder-driven transition.
	The transition occurs at the energy $E=0$ near the (renormalised\cite{Syzranov:unconv}) band edge
	or near a nodal point (for semimetals).}
	\label{PhaseDiagr}
\end{figure}
{\it The phase diagram} of a finite-size system near a non-Anderson disorder-driven transition
is shown in Fig.~\ref{PhaseDiagr}. In what follows we 
measure all energies from a nodal point or a (renormalised\cite{Syzranov:unconv}) band
edge[45] set to $E=0$.

For disorder strengths weaker than a critical value (``weak-disorder phase''
in Fig.~\ref{PhaseDiagr}), $\varkappa<\varkappa_c$, the disorder is
perturbatively irrelevant\cite{Syzranov:Weyl,Syzranov:unconv}
with the dimensionless disorder strength  $\gamma(E)\sim[\tau(E) E]^{-1}$ vanishing at small energies
$E\rightarrow 0$,
where $\tau(E)$ is the elastic scattering time. 

Unlike the case of low dimensions, the lowest-energy levels of a sufficiently large high-dimensional
system are {\it discrete} for $\varkappa<\varkappa_c$, 
as the ``level width'' $\tau(E)^{-1}$ vanishes faster than the spatial-quantisation
gaps $\sim E\propto L^{-\alpha}$ between the lowest levels at $E\rightarrow0$.
The interplay of the level discreteness with multifractality will be discussed below.  

For supercritical disorder strength, $\varkappa>\varkappa_c$,
the dimensionless disorder strength grows at low energies, until reaching
the value $\gamma\sim 1$ that marks the boundary of the ``strong-disorder phase'' (that for $d>2$-dimensional
semiconductors in the orthogonal symmetry class also matches the mobility threshold) in Fig.~\ref{PhaseDiagr}.

{\it Inverse participation ratios.}
The wavefunction structure at energy $E$ and disorder strength $\varkappa$
is conveniently characterised by
the disorder-averaged IPRs\cite{Wegner:IPR,MirlinEvers,Mirlin:review1}
\begin{align}
	&P_q(E,\varkappa)=\frac{\left<\sum_{n}\int|\psi_n(\br)|^{2q}d\br\:\delta(E-E_n)\right>_{\text{dis}}}
	{\left<\sum_{n}\delta(E-E_n)\right>_{\text{dis}}}\equiv
	\nonumber\\
	&\frac{i^{q-2}}{2\pi\rho(E)}\lim_{\eta\rightarrow0}(2\eta)^{q-1}
	\left<\left[G_R(\br,\br,E,\eta)\right]^{q-1}G_A(\br,\br,E,\eta)\right>_{\text{dis}},
	\label{IPRmain}
\end{align}
where $G_{R,A}(\br,\br^\prime,E,\eta)=\sum_n \frac{\psi_n^*(\br)\psi_n(\br)} 
{E-E_n\pm i\eta}$ are the retarded and advanced Green's functions
with an artificially introduced ``dephasing rate'' $\eta$,
$E_n$ are the energies of the eigenstates $\psi_n$ for a given disorder realisation,
and $\rho(E)$ is the (disorder-averaged) density of states.

\begin{figure}[t]
	\centering
	\includegraphics[width=\columnwidth]{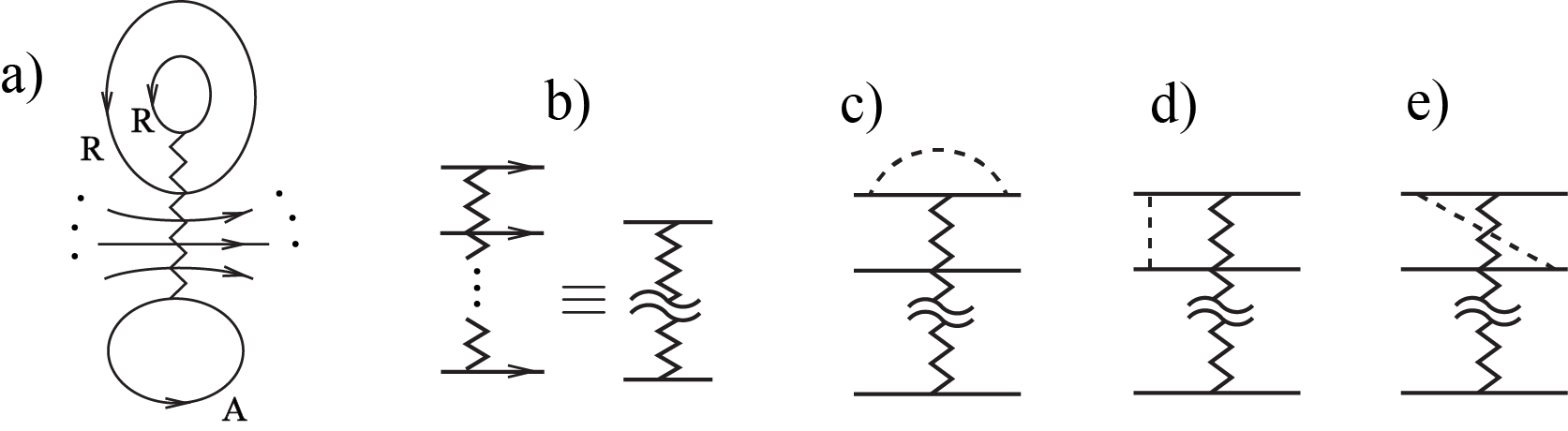}
	\caption{Diagrams for calculating the IPR. a) Product of $q$ Green's functions before disorder
	averaging, cf. Eq.~(\ref{IPRmain}). b) Vertex that connects $q$ Green's functions. c)-e) Diagrams for the one-loop renormalisation
	of the vertex. For $\epsilon_\bk\propto|\bk|^\alpha$, diagrams c)-e) have equal values.
	In Weyl semimetals and other odd-spectrum ($\epsilon_\bk=-\epsilon_{-\bk}$) systems diagrams d) and e) cancel
	each other.
	}
	\label{BasicDiagr}
\end{figure}

Below we compute
the disorder-averaged IPRs (\ref{IPRmain}) near the critical point $(E=0,\varkappa=\varkappa_c)$
as a function of the system size $L$
or the localisation length $\xi$.
The IPRs, Eq.~(\ref{IPRmain}), are mimicked by the diagram in Fig.~\ref{BasicDiagr}a.

Unlike the case of low dimensions, $d<2\alpha$, the low-energy 
properties of high-dimensional materials under consideration
are affected by elastic scattering between all momenta in the band\cite{Syzranov:unconv}.
Such ultraviolet processes are qualitatively important on length scales shorter than
the mean free path $\ell$, and, in particular, determine the criticality and multifractality
near the transition point due to the diverging $\ell$. This leads to the critical properties and
multifractal spectrum different from those at the usual AL transition, that occurs
for states away from nodes and band edges and is described 
by non-linear sigma-models\cite{Efetov:book} on length scales longer than the mean free path.

{\it Renormalisation procedure.}
The effects of the ultraviolet scattering
can be addressed in a controlled way 
by means of a perturbative renormalisation-group (RG) 
controlled by the small parameter\cite{Goswami:TIRG}\cite{Syzranov:Weyl,Syzranov:unconv}
\cite{Pixley:ExactZpubliahed,Syzranov:twoloopZ}
$\varepsilon=2\alpha-d$. The results obtained from this approach are expected to hold qualitatively also
in systems with $|\varepsilon|\gtrsim 1$.

To perform renormalisations, we rewrite the disorder-averaged IPRs using a supersymmetric\cite{Efetov:book}
field theory:
\begin{subequations}
\begin{align}
	P_q
	=&\lim_{\eta\rightarrow0}\frac{(2\eta)^{q-1}}{2\pi\rho(E)}V(K)
	\int{\cal D}\psi {\cal D}\psi^\dagger
	\left[s_R(\br)s_R^*(\br)\right]^{q-1}
	\nonumber\\
	&s_A(\br)s_A^*(\br)
	\:\exp({-\cL_0-\cL_{\text{int}}}),
	\label{PqSuper}
	\\
	\cL_0=&-i\int \psi^\dagger\Lambda^\frac{1}{2}
	\left[E\,\lambda(K)+i\eta\Lambda\lambda(K)-\epsilon_{\hat\bk}\right]
	\Lambda^\frac{1}{2}\psi\,d\br,
	\label{L0}
	\\
	\cL_{\text{int}}&=\frac{1}{2}\varkappa(K)\int(\psi^\dagger\Lambda\psi)^2d\,\br,
	\label{Lint}
\end{align}
\end{subequations}
where 
$\psi^\dagger$ and $\psi$ are $4$-component supervectors
in the $FB\otimes RA$ (fermion-boson $\otimes$ retarded-advanced) space, $s$ and $s^*$ are the bosonic components
of the supervectors, and
$\Lambda=\left(\sigma_z\right)_{RA}\otimes\holOne_{FB}$.
The factors with $\Lambda$-matrices in Eq.~(\ref{L0}) ensure the convergence of the supersymmetric integral
with respect to the bosonic variables\cite{Mirlin:review1}.

Upon repeatedly integrating out shells of highest momenta,
the action (\ref{L0})-(\ref{Lint}) reproduces itself with 
renormalised disorder strength $\varkappa(K)$ and the parameters $\lambda(K)$ and $V(K)$
that ``flow'' with the running cutoff $K$ and initial values $\varkappa(K_0)=\varkappa_0$,
$\lambda(K)=1$, and $V(K)=1$, where $K_0$ is the ultraviolet momentum cutoff set by
the bandwidth or the impurity size\cite{Syzranov:unconv}.

The parameters $\lambda$ and $V$ grow upon coarse-graining and exhibit singular behaviour
with
\begin{align}
	V\propto\lambda^\zeta
	\label{zeta}
\end{align}
at the
critical disorder, $\varkappa=\varkappa_c$.
The IPRs~(\ref{IPRmain}) can be rewritten as
\begin{align}
	P_q(E,\varkappa_0,K_0)=\lambda^{\zeta-q} P_q[\lambda(K)E,\varkappa(K),K],
	\label{PqRenormGen}
\end{align}
where $P_q[\lambda(K)E,\varkappa(K),K]$ is the IPR of an effective renormalised system with the same quasiparticle
dispersion $\epsilon_\bk\propto k^\alpha$, but with renormalised disorder strength $\varkappa$ and energy $\lambda E$
and that excludes scattering into momentum states $k>K$ that were 
removed by the RG procedure.

The RG has to be stopped either when the spatial quantisation effects become important 
or if it runs into the regime of strong disorder, $\varkappa K^{-\varepsilon}\sim 1$.
The renormalised system is then 
equivalent to a simple (low-dimensional)
system with discrete energy levels or with 
a constant density of states
and unaffected by scattering into high-momentum modes; the IPR $P_q$ of such a system can be found
using conventional methods developed for low-dimensional systems\cite{Efetov:book,Mirlin:review1}. 

{\it Fractality of delocalised states.}
In what immediately follows we consider a system with delocalised finite-energy states 
at $\varkappa=\varkappa_c$ [along path $1$ in Fig.~(\ref{PhaseDiagr})], as, e.g., in a $d>\max(2,2\alpha)$-dimensional system with potential disorder.

The RG procedure at critical disorder is terminated 
when either the momentum $K$ reaches $1/L$ or the spatial quantisation effects
become important (i.e. the energy levels become discrete).
The renormalised system is then either ballistic
(for small $\varepsilon$, that ensures weak disorder at the critical point)
or equivalent to a usual weakly disordered metal (for larger $\varepsilon$), 
with\cite{Efetov:book,MirlinEvers} $P_q\propto L^{-d(q-1)}$.

The characteristic energy $E_L$ of terminating the RG is related to the
system size $L$ as $E_L\propto L^{-z}$, where $z$ is the dynamical critical exponent.
Using that $\lambda(K\sim L^{-1})E_L\sim \epsilon_{K\sim L^{-1}}\propto L^{-\alpha}$, we find 
$\lambda(K\sim L^{-1})\propto L^{z-\alpha}$, which, together with Eq.~(\ref{PqRenormGen}) gives the multifractal spectrum
\begin{align}
	\Delta_q=-(\zeta-q)(z-\alpha).
	\label{DeltaGeneral}
\end{align}

{\it Localised states.}
For trivial-topology systems in the orthogonal symmetry class
the states in the ``strong-disorder phase'' (Fig.~\ref{PhaseDiagr}) are localised.
Also, all states on the phase diagram are localised in systems in $d\leq2$ dimensions
(if allowed by symmetry/topology).

Near the critical point of the non-Anderson disorder-driven transition localised states are
still multifractal on length scales shorter than the localisation length $\xi$ (that diverges at the 
transition).

For zero-energy states at supercritical disorder, $\varkappa>\varkappa_c$, (along
path 2 in Fig.~\ref{PhaseDiagr})
the RG is stopped when reaching strong disorder, $\varkappa(K)K^{-\varepsilon}\sim1$. 
Such $E=0$-states are then characterised by only one length scale $K^{-1}$ that
gives the localisation length $\xi$. Similarly to the case of delocalised states
in a size-$L$ system, we find that $\lambda(K\lesssim \xi^{-1})\propto \xi^{z-\alpha}$,
and the participation ratio
\begin{align}
	P_q\propto \xi^{-d(q-1)-\Delta_q}
	\label{IPRloc}
\end{align}
with the multifractal spectrum (\ref{DeltaGeneral}). The non-trivial scaling of the IPR~(\ref{IPRloc})
with the size of the localisation cell reflects the multifractality of the wavefunctions
on length scales $L<\xi$.

{\it Orthogonal semiconductors.} For a disordered system with the quasiparticle dispersion $\epsilon_\bk=a|\bk|^\alpha$
the RG flow of the system parameters is given in terms of the dimensionless disorder strength
$\gamma=4C_d\varkappa K^{-\varepsilon}/a^2$, with $C_d=2^{1-d}\pi^{-\frac{d}{2}}/\Gamma\left(\frac{d}{2}\right)$, by
the RG equations
\begin{subequations}
\begin{align}
	\partial_l V &={q(2q-1)}V\gamma/4+\ldots,
	\label{RGVO}
	\\
	\partial_l \lambda &=\gamma\lambda/4+\ldots,
	\label{RGLambdaO}
	\\
	\partial_l \gamma &= \varepsilon\gamma+\gamma^2+\ldots.
	\label{RGGammaO}
\end{align}
\end{subequations}
where $l=\ln(K_0/K)$ and $\ldots$ are the terms of higher orders in $\gamma$.
Eqs.~(\ref{RGLambdaO}) and (\ref{RGGammaO}) for the renormalisation of the energy
and the disorder strength have been obtained previously in Refs.~\onlinecite{Syzranov:Weyl} and
\onlinecite{Syzranov:unconv}.
Eq.~(\ref{RGVO}) describes the flow of the preexponential in Eq.~(\ref{PqSuper}).

The renormalisations (\ref{RGVO})-(\ref{RGGammaO}) can be also easily obtained diagrammatically.
For instance, the one-loop renormalisation of the vertex $V$, Fig.~\ref{BasicDiagr}b, is given by $q$ diagrams equivalent
to \ref{BasicDiagr}c, $q(q-1)$ diagrams equivalent to \ref{BasicDiagr}d, and $q(q-1)$ diagrams equivalent to
\ref{BasicDiagr}e. All these diagrams have the same value for the dispersion under consideration, hence the
prefactor $q(2q-1)$ in Eq.~(\ref{RGVO}). The renormalisation of the disorder strength $\gamma$ 
and the parameter $\lambda$
for the dispersion under consideration has been described in detail in Ref.~\onlinecite{Syzranov:unconv}. 

In the one-loop order we find from Eqs.~(\ref{RGVO})-(\ref{RGGammaO}) that $\zeta\approx q(2q-1)$
and\cite{Syzranov:unconv} $z\approx \alpha-\varepsilon/4$, which, together with Eq.~(\ref{DeltaGeneral}),
gives the multifractal spectrum (\ref{SemicondResult}).

{\it Chiral systems,} such as Weyl semimetals or chiral chains with long-range hopping\cite{Garttner:longrange}, often have odd quasiparticle spectra, $\epsilon_\bk=-\epsilon_{-\bk}$, which leads to the mutual
cancellation of diagrams \ref{BasicDiagr}d and \ref{BasicDiagr}e. The RG flow of the vertex $V$
is given by Eq.~(\ref{RGVO}) with the replacement $q(2q-1)\rightarrow 2q$ and with the dimensionless
disorder strength $\gamma=2C_d K^{-\varepsilon}\varkappa$. The flow of $\lambda$ is given by Eq.~(\ref{RGLambdaO})
with the replacement $4\rightarrow2$.
From the RG equations we find $\zeta=q+\cO(\varepsilon)$ [cf. Eq.~(\ref{zeta})], which, according to
Eq.~\eqref{DeltaGeneral}, gives vanishing multifractality $\Delta_q=0$ in the one-loop order.
Thus, finding the multifractal behaviour in such chiral systems requires RG analysis in higher
orders.

In what follows we present the result for a Weyl semimetal (see Appendix
for a detailed two-loop RG analysis of multifractality using the minimal-subtraction scheme).

{\it Weyl semimetals} are 3D systems with the quasiparticle dispersion $\epsilon_{\bk}=\hbsigma\bk$,
where $\hbsigma$ is a vector of Pauli matrices.
WSM properties near the non-Anderson disorder-driven transition can be studied by performing RG
analysis in $d=2-\varepsilon$ dimensions with setting $\varepsilon=-1$ at the end of the calculation.
Although the RG procedure is controlled by small $\varepsilon$, it
is known\cite{Brouwer:WSMcond,Garttner:longrange,LiuOhtsuki:LateNumerics,Pixley:ExactZpubliahed,Bera:Weyl}
to give good agreement with numerical results 
even for $\varepsilon\sim1$. 

The flows of the parameters of a disordered WSM in $2-\varepsilon$ dimensions
are given by (see Appendix)
\begin{subequations}
\begin{align}
	\partial_l V &=\frac{q}{2}V\gamma+\left(\frac{3}{8}q^2-\frac{q}{4}\right)V\gamma^2+\ldots,
	\label{VRGWSM}	
	\\
	\partial_l \lambda &=\frac{1}{2}\lambda\gamma+\frac{1}{8}\lambda\gamma^2+\ldots,
	\label{LambdaRGWSM}
	\\
	\partial_l \gamma &= \varepsilon\gamma+\gamma^2+\frac{1}{2}\gamma^3+\ldots.
	\label{GammaRGWSM}
\end{align}
\end{subequations}
Eqs.~(\ref{LambdaRGWSM}) and (\ref{GammaRGWSM}) for the renormalisation of the energy and disorder strength
in a disordered WSM have been derived previously in Refs.~\onlinecite{Ostrovsky:grapheneRGM} and
\onlinecite{Syzranov:twoloopZ} and in Refs.~\onlinecite{Wetzel:twoloopM,Ludwig:twoloopM,Rossi1M,Rossi2M,TracasVlachosM,Ludwig:ThirringM} for the equivalent Gross-Neveu model.
{An equation equivalent to Eq.~(\ref{VRGWSM}) has also been derived in Ref.~\cite{Foster:multifract2D} to describe
the wavefunctions of 2D Dirac fermions on the surface of a 3D topological
insulator\cite{Foster:multifract2D,Foster:multifract2Dnumerics}.}


From Eqs.~(\ref{VRGWSM})-(\ref{GammaRGWSM}) we find to the two-loop order $\zeta=q-\frac{3}{4}(q^2-q)\varepsilon
+\cO(\varepsilon^2)$, which, together with $z=1-\frac{\varepsilon}{2}+\ldots$ and Eq.~(\ref{DeltaGeneral}),
gives the multifractal spectrum (\ref{WeylResult}).

{\it Level discreteness and observability of multifractality.}
Observation of multifractality at the critical point ($E=0$, $\varkappa=\varkappa_c$)
requires that the disorder-averaged energy spectrum of the system at this point is continuous,
i.e. smeared by disorder and unaffected by the spatial quantisation.
This condition is always met in systems with $|\varepsilon|\sim 1$ 
as the ``level width'' 
$\tau(E_n)^{-1}\sim E_n$ for the lowest levels $n$ is of the order of their energies $E_n$.

However, for some systems, e.g., chains of ultracold ions\cite{Monroe:longrange},
it is possible to realise\cite{Garttner:longrange} $|\varepsilon|\ll1$,
that corresponds to weak disorder $\gamma_c\sim |\varepsilon|$ at the critical point
and the existence of a large number $\sim 1/\varepsilon$ of energy levels that remain discrete
[$\tau^{-1}(E_n)\ll|E_n-E_{n-1}|$] at the critical disorder,
although the energies and the spacings between these levels vanish in the limit $L\rightarrow\infty$. 
The multifractal behaviour in such systems is observable only
on sufficiently short length scales
\begin{align}
L<L_{\text{discr}}=L\,|\varepsilon|^\frac{1}{d}
\label{DiscrCond}
\end{align}
that correspond to the wavelengths of higher levels belonging to the continuous part of the energy spectrum.

{\it Rare-region effects.} The perturbative RG that we used in this paper neglects non-perturbative
instantonic contributions[\onlinecite{Suslov:rare},\onlinecite{Syzranov:unconv}] to the field theory~(\ref{PqSuper})-(\ref{Lint}),
that, e.g., result in the
formation of Lifhsitz tails near band edges and always lead to a finite density of states\cite{Wegner:DoS,Nandkishore:rare}
near nodes.
The effects of such instantons on physical observables, such as the density of states and conductivity, are rather small
near the critical point in high-dimensional systems  and were undetectable in all numerical studies\cite{Herbut,Brouwer:WSMcond,Pixley:twotrans,Pixley:ExactZpubliahed,LiuOhtsuki:LateNumerics,Brouwer:exponents,
Shapourian:PhaseDiagr,Bera:Weyl} so far except Ref.~\onlinecite{PixleyHuse:rare}. 

Another potential consequence of rare-region (instantonic) effects,
albeit currently not demonstrated analytically, may be the ``rounding'' of the criticality, i.e. preventing the divergence of the correlation length near the critical point, and thus converting 
the phase transition of the type discussed here into a sharp crossover, with the latter scenario advocated in Ref.~\onlinecite{PixleyHuse:rare}. In our view, the plausibility of this scenario
deserves further investigation, in particular, in systems that disallow localisation by symmetry and topology.
In the case the criticality does get smeared in a system, the wavefunction multifractality studied here is
observable on distances shorter than a large characteristic length set by the rare-region effects.

{\it Acknowledgements.}
We are grateful to A.W.W.~Ludwig for useful motivating discussions at the initial stages of
this work and to P.M.~Ostrovsky for previous collaboration.
{Also, we thank M.S. Foster for valuable discussions.}
Our work was financially supported by the NSF grants
DMR-1001240 (LR and SVS), DMR-1205303(VG and SVS), PHY-1211914 (VG and SVS), PHY-1125844 (SVS) and
by the Simons Investigator award of the Simons Foundation (LR).


\newpage
\appendix
\section{Appendix: Two-loop RG flow in a Weyl semimetal}

Here we present a two-loop renormalisation-group analysis of the multifractal properties
of a disordered Weyl semimetal in $d=2-\varepsilon$ dimensions using the minimal subtraction 
scheme\cite{PeskinSchroeder}. 

A similar scheme has been applied recently
\cite{Syzranov:twoloopZ} to compute the correlation-length
and dynamical exponents for a WSM to the two-loop order,
following previous studies of graphene\cite{Ostrovsky:grapheneRGM}
and of the equivalent Gross-Neveu model\cite{Wetzel:twoloopM,Ludwig:twoloopM,Rossi1M,Rossi2M,TracasVlachosM,Ludwig:ThirringM}.
{Also, a similar multifractality analysis has been 
carried out in Ref.~\cite{Foster:multifract2D} for 2D ($\varepsilon=0$) Dirac fermions on the surface of
a 3D topological insulator. Although there is no disorder-driven phase transition in such a 2D system,
the wavefunctions display multifractal behaviour on sufficiently short length scales.}

\subsection*{Renormalisation scheme}

For performing two-loop renormalisation-group analysis
it is convenient to rewrite the IPRs~(4)
as a derivative of a partition function with respect to
a supersymmetry-breaking term:
\begin{subequations}
\begin{align}
	P_q
	=&-\lim_{\eta\rightarrow0}\frac{(2\eta)^{q-1}}{2\pi\rho(E)}
	\partial_{V_0} Z \vert_{V_0=0}
	\label{IPRminimalS}
	\\
	Z=
	&\int{\cal D}\psi {\cal D}\psi^\dagger \:\exp({-\cL_0-\cL_{\text{int}}-\cL_{V_0}})
	\label{Zpartition}
	\\	
	\cL_0=&-i\int \psi^\dagger
	\left[i\omega+i\eta\Lambda-\hbsigma\hat\bk\right]
	\psi\,d\br,
	\label{L0minimal}
	\\
	\cL_{\text{int}}=&\frac{1}{2}\varkappa_0\int(\psi^\dagger\psi)^2d\,\br
	\\
	\cL_{V_0}=
	&V_0\int\left[s_R(\br)s_R^*(\br)\right]^{q-1} s_A(\br)s_A^*(\br) d\br.
	\label{LV0}
\end{align}
\end{subequations}
Here we have introduced a positive Matsubara frequency $\omega>0$ that ensures the convergence of the
superintegral (\ref{Zpartition})
with respect to the bosonic components of the supervectors $\psi$ and
$\psi^\dagger$ and, in the below calculation, also regularises infrared divergences of momentum integrals
in $d=2-\varepsilon$ dimensions for $\varepsilon>0$.

The Lagrangian of the system in the minimal subtraction scheme is separated into the effective
Lagrangian $\cL_E$ that describes the long-wave behaviour of the physical observables and the
counterterm Lagrangian $\cL_{\text{counter}}$ that cancels contributions divergent in the powers of $1/\varepsilon$:
\begin{subequations}
\begin{align}
	\cL=&\cL_E+\cL_{\text{counter}},
	\label{Lfull}
	\\
	\cL_E=&-i\int \Psi^\dagger
	\left[i\Omega+i\eta\lambda\Lambda-\hbsigma\hat\bk\right]
	\Psi\,d\br
	\nonumber
	\\
	&+\frac{1}{2}\varkappa\int(\Psi^\dagger\Psi)^2d\,\br
	\label{LE}
	\\
	&+V\int\left[S_R(\br)S_R^*(\br)\right]^{q-1} S_A(\br)S_A^*(\br) d\br,
	\label{LV}
\end{align}
\end{subequations}
where the velocity of the renormalised Weyl fermions (the coefficient before $\hbsigma\hat\bk$)
is set to unity, without loss of generality, by appropriately choosing the field $\Psi$.
The scale $\Omega>0$ sets the characteristic momentum of the long-wave behaviour.
The coefficients $V$ and $V_0$ before the source terms (\ref{LV0})
and (\ref{LV}) are considered infinitesimal in the calculation below.

We note, that in general the renormalisation generates additional terms
$\propto V\int\left[S_R(\br)S_R^*(\br)\right]^{m} S_A(\br)S_A^*(\br) d\br $ with $m<q-1$,
which we neglect here because their contributions to the partition function
(\ref{Zpartition})
are less singular $\propto\eta^{-(m-1)}$ at $\eta\rightarrow0$ than that of the term (\ref{LV}), and, thus,
they do not contribute to the IPRs~(\ref{IPRminimalS}).

The minimal subtraction scheme\cite{PeskinSchroeder} consists in
calculating perturbative corrections to the Lagrangian (\ref{LE})
and choosing the counterterms $\cL_{\text{counter}}$
to cancel divergent in powers of $1/\varepsilon$ contributions. The RG equations can then
be derived by relating the ``observable''
parameters $\Psi$, $\Omega$, $\varkappa$, and $V$ to the
``bare'' ones $\psi$, $\omega$, $\varkappa_0$, and $V_0$.

All momentum integrals in such a calculation are convergent in low
dimensions $d=2-\varepsilon$ with $\varepsilon>0$.
The results have to be analytically continued to higher dimensions, $\varepsilon<0$,
at the end of the calculation (dimensional regularisation).
The ``dephasing rate'' $\eta$, sent to zero at the end of the calculation 
[cf. Eqs.~(\ref{IPRminimalS}) and (\ref{L0minimal})],
may be assumed to be significantly smaller than the scale $\Omega$ and neglected when computing
the parameters of the renormalised Lagrangian.

\begin{widetext}
\begin{figure*}
	
	\includegraphics[width=\textwidth]{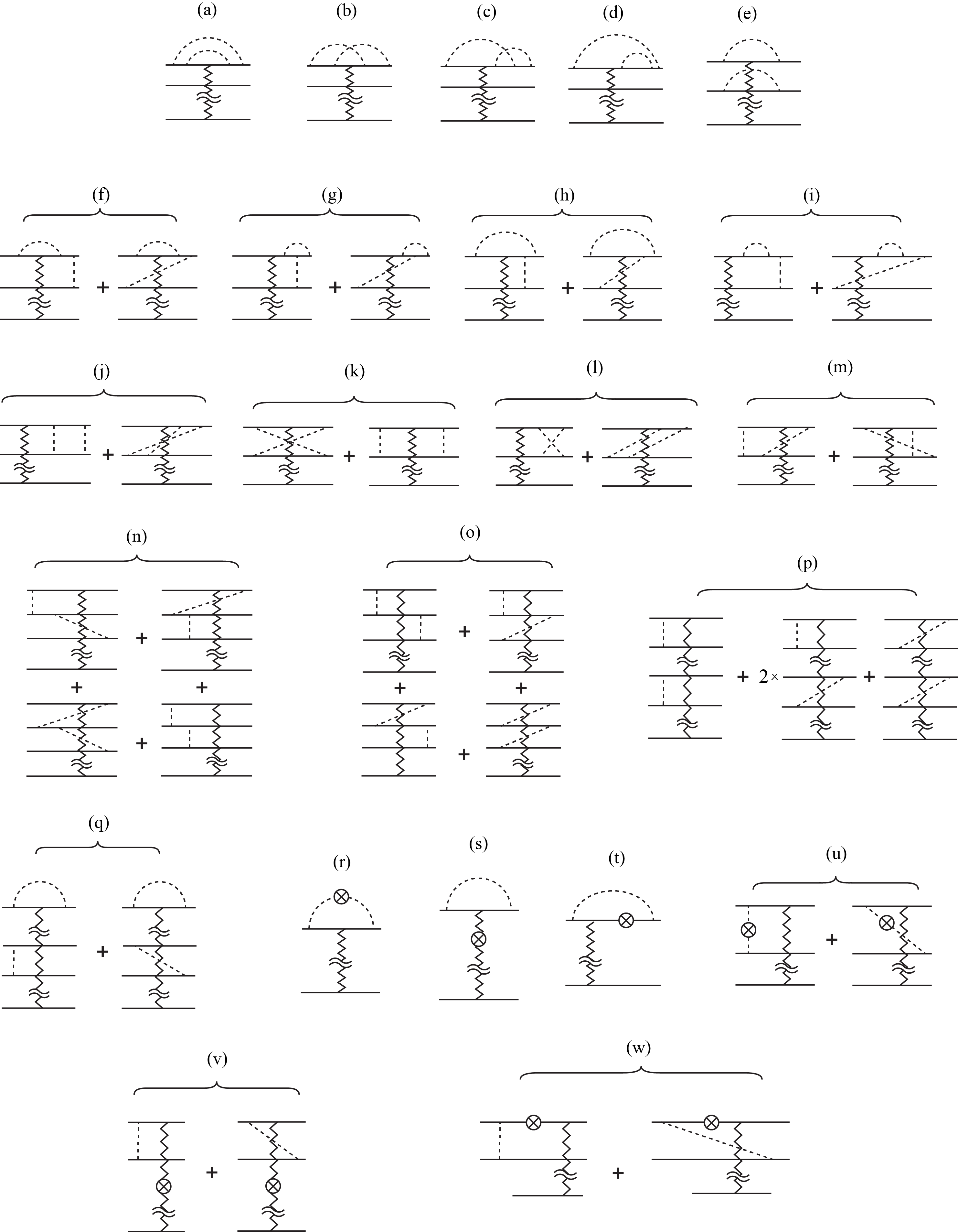}
	\caption{Diagrams for the two-loop renormalisation of the vertex $V$.}
	\label{Diagrams}
\end{figure*}

\end{widetext}

\subsection*{One-loop renormalisations}

In the one-loop order the renormalisation of the disorder strength $\varkappa$ and
the quasiparticle energy $\Omega$ has
been considered in detail in Ref.~\onlinecite{Syzranov:twoloopZ}.

The one-loop perturbative correction to the vertex $V$ is described by the diagrams in Fig.~2c-e
and the topologically equivalent diagrams. Since the ``dephasing rate'' $\eta$ may be neglected
when considering the respective high-momentum scattering processes, the advanced and retarded Green's
functions may be taken identical $G^A(\bp,i\Omega)=G^R(\bp,i\Omega)=(i\Omega-\hbsigma\bp)^{-1}$
when evaluating these diagrams.
The sum of diagrams~2c-e is given by
\begin{align}
	&(\delta V)^{\text{1-Loop}}
	\nonumber\\
	&=q(q-1)V\varkappa\int_\bp\frac{1}{i\Omega-\hbsigma\bp}\otimes
	\left(\frac{1}{i\Omega+\hbsigma\bp}+\frac{1}{i\Omega-\hbsigma\bp}\right)
	\nonumber\\
	&+qV\varkappa\int_\bp\frac{1}{(i\Omega-\hbsigma\bp)^2}
	=-\frac{q}{\varepsilon}\varkappa V C_{2-\varepsilon}\Omega^{-\varepsilon}+\cO(1),
\end{align}
where the prefactors $q(q-1)$ and $q$ account for the numbers of topologically equivalent diagrams,
$\int_\bp\ldots=\int d^d\bp/(2\pi)^d\ldots$, and $C_d=2^{1-d}\pi^{-\frac{d}{2}}/\Gamma\left(\frac{d}{2}\right)$,
and $\ldots\otimes\ldots$ is the tensor product of the two subspaces of the two $2\times2$ propagators
connected by impurity lines in Figs.~2d and 2e; the structure of the 
correction is trivial in the other propagators' subspaces.

The perturbative corrections to the disorder strength $\varkappa$ and the frequency 
$\Omega$ have been calculated in
detail in Ref.~\onlinecite{Syzranov:twoloopZ}. The velocity of the Weyl fermions [the coefficient before the 
$\hbsigma\hat\bk$ term in Eq.~(\ref{LE})] does not receive first-order corrections.

The singular part $\propto 1/\varepsilon$ of the
one-loop corrections to the Lagrangian is cancelled by the counterterms
\begin{align}
	\cL_{\text{counter}}^{(1)}=&\int \Psi^\dagger\: \delta^{(1)}\Omega\: \Psi\, d\br
	+\frac{1}{2}\delta^{(1)}\varkappa\int (\Psi^\dagger\Psi)^2\,d\br
	\nonumber\\
	&+\delta^{(1)}V\int\left[S_R(\br)S_R^*(\br)\right]^{q-1} S_A(\br)S_A^*(\br)\, d\br
	\label{Lcounter1}
\end{align}
with
\begin{subequations}
\begin{align}
	\delta^{(1)}\Omega= &-\frac{1}{\varepsilon}\Omega\cdot \varkappa C_{2-\varepsilon}\Omega^{-\varepsilon},
	\\
	\delta^{(1)}\varkappa= &-\frac{2}{\varepsilon}\varkappa^2C_{2-\varepsilon}\Omega^{-\varepsilon},
	\\
	\delta^{(1)}V=&
	-\frac{q}{\varepsilon}V\varkappa C_{2-\varepsilon}\Omega^{-\varepsilon}.
	\label{V1loopCounter}
\end{align}
\end{subequations}

\begin{figure}[htbp]
	\centering
	\includegraphics[width=0.7\columnwidth]{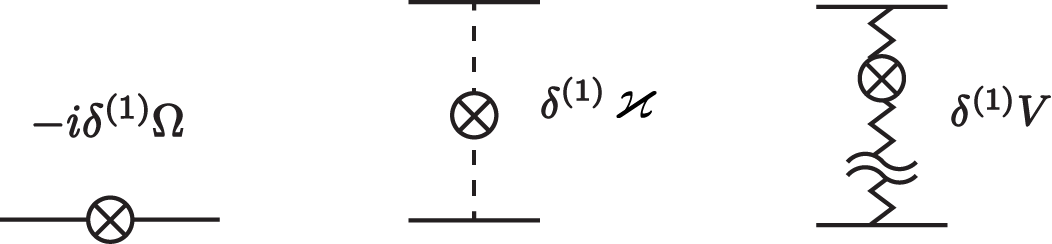}
	\caption{Additional diagrammatic elements that come from the one-loop counterterms.}
	\label{OneLoopCounter}
\end{figure}

The presence of the one-loop counterterms (\ref{Lcounter1})
requires introducing the respective diagrammatic elements, Fig.~\ref{OneLoopCounter},
in addition to the propagators and impurity lines,
when calculating diagrammatically perturbative corrections beyond the one-loop order.

\subsection*{Two-loop diagrams for the source-term renormalisation}

The diagrams that describe the renormalisation of the source term $\propto V$ 
in the two-loop order are shown in Fig.~\ref{Diagrams}.

Diagrams in Fig.~\ref{Diagrams}(a)-(m) are computed similarly to the two-loop diagrams for the
renormalisation of the disorder strength, obtained by replacing the zigzag line by an impurity line and
considered in detail in Ref.~\onlinecite{Syzranov:twoloopZ}.
The values of diagrams \ref{Diagrams}(a)-(m), together with the numbers of equivalent diagrams,
are provided in Table~\ref{Table:SameDiagr}.

\begin{figure}[h!]
	\centering
	\includegraphics[width=0.85\columnwidth]{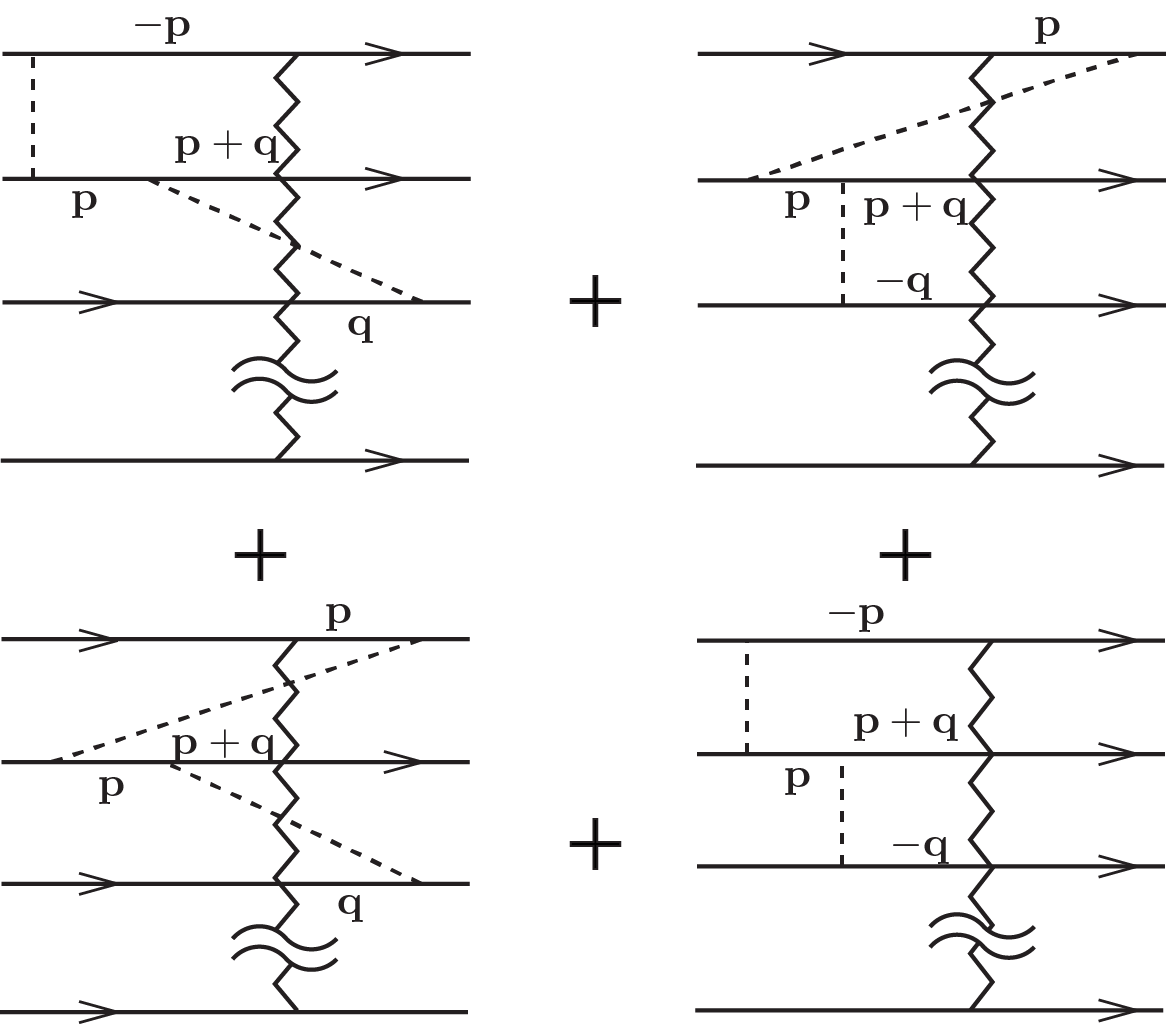}
	\caption{Momenta in diagram (n) in Fig.~\ref{Diagrams}.
}
	\label{DiagramN}
\end{figure}

Diagrams \ref{Diagrams}(n)-(p) are regular in $\varepsilon$ due to the mutual cancellation
of the singularities coming from blocks with vertical and diagonal impurity lines.
For instance, the sum of the diagrams in Fig.~\ref{Diagrams}(n) (see also Fig.~\ref{DiagramN}) can be evaluated
(in units $V\varkappa^2$) as
\begin{align}
	\int\frac{1}{i\Omega-\hbsigma\bp}\frac{1}{i\Omega-\hbsigma(\bp+\bq)}
	\otimes
	\left(\frac{1}{i\Omega-\hbsigma\bq}+\frac{1}{i\Omega+\hbsigma\bq}\right)
	\nonumber\\
	\otimes
	\left(\frac{1}{i\Omega-\hbsigma\bp}+\frac{1}{i\Omega+\hbsigma\bp}\right)
	\nonumber\\
	=4\Omega^2\int\frac{(i\Omega+\hbsigma\bp)[i\Omega+\hbsigma(\bp+\bq)]}
	{(\Omega^2+p^2)^2(\Omega^2+q^2)[\Omega^2+(\bp+\bq)^2]}
	\nonumber\\
	=4\Omega^2\int\frac{(\hbsigma\bp)(\hbsigma\bq)}
	{(\Omega^2+p^2)^2(\Omega^2+q^2)[\Omega^2+(\bp+\bq)^2]}
	\nonumber\\
	+4\Omega^2\int\frac{1}
	{(\Omega^2+p^2)(\Omega^2+q^2)[\Omega^2+(\bp+\bq)^2]}
	\nonumber\\
	-8\Omega^4\int\frac{1}
	{(\Omega^2+p^2)^2(\Omega^2+q^2)[\Omega^2+(\bp+\bq)^2]}
	\nonumber\\
	=\cO(1).
\end{align}
(For detailed calculations of the last integrals see Ref.~\onlinecite{Syzranov:twoloopZ}).
The values of diagrams 3(n)-(q) are given in Table~\ref{DiagrNQ}.

Similarly one can show the vanishing of 
singular in $1/\varepsilon$ contributions in
diagrams \ref{Diagrams}(o) and \ref{Diagrams}(p).

Diagrams \ref{Diagrams}(q) are given by (in units $V\varkappa^2$)
\begin{align}
	\int_\bq\frac{1}{(i\Omega-\hbsigma\bq)^2}
	\otimes
	\int_\bp\frac{1}{i\Omega-\hbsigma\bp}
	\nonumber\\
	\otimes
	\left(
	\frac{1}{i\Omega-\hbsigma\bp}
	+\frac{1}{i\Omega+\hbsigma\bp}
	\right)
	\nonumber\\
	=-\frac{1}{\varepsilon}\left(C_{2-\varepsilon}\Omega^{-\varepsilon}\right)^2
	+\cO(1)
\end{align}

Diagrams \ref{Diagrams}(r)-(w) are the two-loop diagrams for the
corrections to the vertex $V$ that contain one-loop counterterms
and are equivalent to similar diagrams in Ref.~\onlinecite{Syzranov:twoloopZ}, up
to replacing the zigzag line or its counterterm by the impurity line or its counterterm,
with the values provided in Table~\ref{Table:Counter}.

\begin{table}
\begin{tabular}{c|c|c}
Diagram & \# Equivalent diagrams & Value \\ \hline\hline
(a) & $q$ & $\frac{1}{\varepsilon^2}-\frac{2}{\varepsilon}$ \vspace{1mm}\\ \hline
(b) & $q$ & $-\frac{1}{2\varepsilon^2}+\frac{2}{\varepsilon}$ \vspace{1mm} \\ \hline
(c) & $2q$ & $\frac{1}{2\varepsilon^2}-\frac{1}{\varepsilon}$ \vspace{1mm} \\ \hline
(d) & $2q$ & $-\frac{1}{2\varepsilon}$ \vspace{1mm} \\ \hline
(e) & $\frac{q(q-1)}{2}$ & $\frac{1}{\varepsilon^2}-\frac{2}{\varepsilon}$ \vspace{1mm} \\ \hline
(f) & $2q(q-1)$ & $-\frac{1}{\varepsilon}$ \vspace{1mm} \\ \hline
(g) & $2q(q-1)$ & $-\frac{1}{\varepsilon}$ \vspace{1mm} \\ \hline
(h) & $2q(q-1)$ & 0 \\ \hline
(i) & $2q(q-1)$ & 0 \\ \hline
(j) & $q(q-1)$ & $\frac{1}{\varepsilon^2}-\frac{1}{2\varepsilon}$ \vspace{1mm} \\ \hline
(k) & $\frac{q(q-1)}{2}$ & $\frac{1}{\varepsilon^2}-\frac{1}{2\varepsilon}$ \vspace{1mm} \\ \hline
(l) & ${q(q-1)}$ & $-\frac{1}{2\varepsilon^2}+\frac{1}{2\varepsilon}$ \vspace{1mm} \\ \hline
(m) & $2{q(q-1)}$ & $-\frac{1}{2\varepsilon^2}+\frac{1}{2\varepsilon}$ \vspace{1mm} \\ \hline\hline
\end{tabular}
\caption{
\label{Table:SameDiagr}
Numbers and values [in units $V\left(C_d\Omega^{-\varepsilon}\varkappa\right)^2$,
up to $\cO(1/\varepsilon)$]
of the diagrams equivalent to (a)-(m) in Fig.~\ref{Diagrams}.}
\end{table}

\begin{table}
\begin{tabular}{c|c|c}
Diagram & \# Equivalent diagrams & Value \\ \hline\hline
(n) & $2q(q-1)(q-2)$ & $0$ \\ \hline
(o) & $q(q-1)(q-2)$ & $0$ \\ \hline
(p) & $\frac{q(q-1)(q-2)(q-3)}{2}$ & $0$ \vspace{1mm} \\ \hline
(q) & $q(q-1)(q-2)$ & $-\frac{1}{\varepsilon}$ \vspace{1mm} \\ \hline
\end{tabular}
\caption{
\label{DiagrNQ}
Numbers and values [in units $V\left(C_d\Omega^{-\varepsilon}\varkappa\right)^2$,
up to $\cO(1/\varepsilon)$] 
of the diagrams equivalent to (n)-(q) in Fig.~\ref{Diagrams}. These diagrams exist only for $q>2$.}
\end{table}

\begin{table}
\begin{tabular}{c|c|c}
Diagram & \# Equivalent diagrams & Value \\ \hline\hline
(r) & $q$ & $-\frac{2}{\varepsilon^2}+\frac{2}{\varepsilon}$ \vspace{1mm} \\ \hline
(s) & $q$ & $-\frac{q}{\varepsilon^2}+\frac{q}{\varepsilon}$ \vspace{1mm} \\ \hline
(t) & $2q$ & $\frac{1}{2\varepsilon}$ \vspace{1mm} \\ \hline
(u) & $q(q-1)$ & $\frac{2}{\varepsilon}$ \vspace{1mm} \\ \hline
(v) & $q(q-1)$ & $\frac{q}{\varepsilon}$ \vspace{1mm} \\ \hline
(w) & $2q(q-1)$ & $0$ \\ \hline
\end{tabular}
\caption{
\label{Table:Counter}
Numbers and values [in units $V\left(C_d\Omega^{-\varepsilon}\varkappa\right)^2$,
up to $\cO(1/\varepsilon)$]
of the diagrams [(r)-(w) in Fig.~\ref{Diagrams}] for the two-loop renormalisation
of the vertex $V$ that include one-loop counterterms.}
\end{table}

\subsection*{RG equations}

The one-loop and two-loop corrections to the Lagrangian~(\ref{LE}) are cancelled by the counterterm
Lagrangian
\begin{align}
	\cL_{\text{counter}} =&-i\int \Psi^\dagger	\left[\delta(i\Omega)-\delta(\hbsigma\hat\bk)\right]\Psi\: d\br
	\nonumber\\
	&+\frac{1}{2}\delta\varkappa\int\left(\Psi^\dagger\Psi\right)^2d\br
	\nonumber\\
	&+\delta V\int\left[S_R(\br)S_R^*(\br)\right]^{q-1} S_A(\br)S_A^*(\br) d\br,
\end{align}
where the values of $\delta(i\Omega)$, $\delta(\hbsigma\hat\bk)$,
and $\delta\varkappa$ have been calculated previously\cite{Syzranov:twoloopZ,Ostrovsky:grapheneRGM}
(see also Refs.~\onlinecite{Wetzel:twoloopM,Ludwig:twoloopM,Rossi1M,Rossi2M,TracasVlachosM,Ludwig:ThirringM}):
\begin{subequations}
\begin{align}
	&\delta(i\Omega)=
	i\Omega\left[-\frac{1}{\varepsilon}\varkappa C_{2-\varepsilon}\Omega^{-\varepsilon}
	+\frac{3}{2}\frac{1}{\varepsilon^2}
	\left(\varkappa C_{2-\varepsilon}\Omega^{-\varepsilon}\right)^2
	\right],
	\label{OmegaDivergent}
	\\
	&\delta(\hbsigma\hat\bk)
	=\frac{1}{4\varepsilon}(\varkappa C_{2-\varepsilon}\Omega^{-\varepsilon})^2\hbsigma\hat\bk,
	\\
	&\delta\varkappa
	=\varkappa\left[-\frac{2}{\varepsilon}\varkappa C_{2-\varepsilon}\Omega^{-\varepsilon}
	+\left(\frac{4}{\varepsilon^2}-\frac{1}{2\varepsilon}\right)\left(\varkappa C_{2-\varepsilon}\Omega^{-\varepsilon}\right)^2\right].
	\label{KappaDivergent}
\end{align}

The counterterm for the vertex $V$ is given by the one-loop contribution~(\ref{V1loopCounter})
minus the sum of the diagrams in Fig.~\ref{Diagrams}:
\begin{align}
	\delta^{(1)}V
	=-\frac{q}{\varepsilon}V\varkappa C_{2-\varepsilon}\Omega^{-\varepsilon}
	+\frac{1}{\varepsilon^2}\left(\frac{q^2}{2}+q\right)V
	\left(\varkappa C_{2-\varepsilon}\Omega^{-\varepsilon}\right)^2
	\nonumber\\
	-\frac{1}{\varepsilon}\left(\frac{3}{4}q^2-\frac{3}{4}q\right)V
	\left(\varkappa C_{2-\varepsilon}\Omega^{-\varepsilon}\right)^2.
\end{align}
\end{subequations}

Introducing the dimensionless disorder strength
\begin{align}
	\gamma=2\varkappa C_{2-\varepsilon}\Omega^{-\varepsilon},
\end{align}
the full Lagrangian (\ref{Lfull}) can be rewritten as
\begin{align}
	\cL=-i\int\Psi^\dagger
	\left[i\Omega\left(1-\frac{\gamma}{2\varepsilon}
	+\frac{3}{8}\frac{\gamma^2}{\varepsilon^2}\right)\right.
	\left.-\hbsigma\hat\bk
	\left(1+\frac{\gamma^2}{16\varepsilon}\right)
	\right]\Psi \,d\br
	\nonumber\\
	+\frac{\gamma\Omega^\varepsilon}{4C_{2-\varepsilon}}
	\left(1-\frac{\gamma}{\varepsilon}-\frac{\gamma^2}{8\varepsilon}
	+\frac{\gamma^2}{\varepsilon^2}\right)\int(\Psi^\dagger\Psi)^2d\br
	\nonumber\\
	+V\left[1-\frac{q}{2\varepsilon}\gamma
	+\frac{1}{\varepsilon^2}\left(\frac{q^2}{8}+\frac{q}{4}\right)\gamma^2
	-\frac{1}{\varepsilon}\left(\frac{3}{16}q^2-\frac{3}{16}q\right)\gamma^2\right]
	\nonumber\\
	\int\left[S_R(\br)S_R^*(\br)\right]^{q-1} S_A(\br)S_A^*(\br) \,d\br.
	\label{LagrangianFinal}
\end{align}

By comparing the Lagrangian (\ref{LagrangianFinal}), that depends on the renormalised
observables $\Psi$, $\Omega$, $\varkappa$, and $V$, with the Lagrangian (\ref{L0minimal})-(\ref{LV0})
expressed in terms of the ``bare'' variables $\psi$, $\omega$, $\varkappa_0$, and $V_0$,
we can relate the ``bare'' and the renormalised observables:
\begin{subequations}
\begin{align}
	Z &= 1+\frac{\gamma^2}{16\varepsilon},\\
	\varkappa_0 &= \frac{\Omega^\varepsilon}{2C_{2-\varepsilon}}\gamma
	\left(1-\frac{\gamma}{\varepsilon}+\frac{\gamma^2}{\varepsilon^2}-\frac{\gamma^2}{4\varepsilon}\right),
	\label{KappaOGamma}
	\\
	\omega &=\Omega\left(1-\frac{\gamma}{2\varepsilon}+\frac{3\gamma^2}{8\varepsilon^2}-\frac{\gamma^2}{16\varepsilon}\right),
	\label{OmegaBare}
	\\
	V_0 &=V
	\left[
	1-\frac{q}{2\varepsilon}\gamma
	+\frac{1}{\varepsilon^2}\left(\frac{q^2}{8}+\frac{q}{4}\right)\gamma^2
	+\frac{1}{\varepsilon}\left(-\frac{3q^2}{16}+\frac{q}{8}\right)\gamma^2
	\right]
\end{align}
\end{subequations}
where $Z$ describes the wavefunction rescaling: $\psi=\Psi Z^\frac{1}{2}$.

The input parameters $V_0$ and $\varkappa_0$ of the Lagrangian are independent of the characteristic
momentum scale $K=\Omega$ at which the long-wave properties of the system are observed, which gives
\begin{align}
	\frac{\partial\ln(\varkappa,V_0)}{\partial\Omega}=0.
	\label{CS}
\end{align}
Eqs.~(\ref{CS}), analogous to the Callan-Symanzik equation\cite{PeskinSchroeder}, 
immediately lead to the RG equations (11c) and (11a) with $l=\ln(K_0/\Omega)$.
The RG equation (11b) follows from Eq.~(\ref{OmegaBare}) using that $\Omega=\lambda\omega$.




\end{document}